\documentclass[aps,notitlepage,singlecolumn,superscriptaddress,superscriptaddress,amsthm,amsmath,amssymb]{revtex4-1}  
\usepackage{graphicx}
\usepackage{dcolumn}
\usepackage{bm}
\usepackage{hyperref}
\usepackage{mathtools}
\usepackage{color}

\usepackage{amssymb,amsmath,amsthm,amsfonts}

\usepackage{xy}
\xyoption{matrix}
\xyoption{frame}
\xyoption{arrow}
\xyoption{arc}

\usepackage{ifpdf}
\ifpdf
\else
\PackageWarningNoLine{Qcircuit}{Qcircuit is loading in Postscript mode.  The Xy-pic options ps and dvips will be loaded.  If you wish to use other Postscript drivers for Xy-pic, you must modify the code in Qcircuit.tex}
\xyoption{ps}
\xyoption{dvips}
\fi

\entrymodifiers={!C\entrybox}

\newcommand{\bra}[1]{{\left\langle{#1}\right\vert}}
\newcommand{\ket}[1]{{\left\vert{#1}\right\rangle}}
\newcommand{\qw}[1][-1]{\ar @{-} [0,#1]}
\newcommand{\qwx}[1][-1]{\ar @{-} [#1,0]}

\newcommand{\cwx}[1][-1]{\ar @{=} [#1,0]}
\newcommand{\gate}[1]{*+<.6em>{#1} \POS ="i","i"+UR;"i"+UL **\dir{-};"i"+DL **\dir{-};"i"+DR **\dir{-};"i"+UR **\dir{-},"i" \qw}
\newcommand{\meter}{*=<1.8em,1.4em>{\xy ="j","j"-<.778em,.322em>;{"j"+<.778em,-.322em> \ellipse ur,_{}},"j"-<0em,.4em>;p+<.5em,.9em> **\dir{-},"j"+<2.2em,2.2em>*{},"j"-<2.2em,2.2em>*{} \endxy} \POS ="i","i"+UR;"i"+UL **\dir{-};"i"+DL **\dir{-};"i"+DR **\dir{-};"i"+UR **\dir{-},"i" \qw}

\newcommand{\control}{*!<0em,.025em>-=-<.2em>{\bullet}}

\newcommand{\ctrl}[1]{\control \qwx[#1] \qw}

\newcommand{\targ}{*+<.02em,.02em>{\xy ="i","i"-<.39em,0em>;"i"+<.39em,0em> **\dir{-}, "i"-<0em,.39em>;"i"+<0em,.39em> **\dir{-},"i"*\xycircle<.4em>{} \endxy} \qw}

\newcommand{\multigate}[2]{*+<1em,.9em>{\hphantom{#2}} \POS [0,0]="i",[0,0].[#1,0]="e",!C *{#2},"e"+UR;"e"+UL **\dir{-};"e"+DL **\dir{-};"e"+DR **\dir{-};"e"+UR **\dir{-},"i" \qw}
\newcommand{\ghost}[1]{*+<1em,.9em>{\hphantom{#1}} \qw}

\newcommand{\gategroup}[6]{\POS"#1,#2"."#3,#2"."#1,#4"."#3,#4"!C*+<#5>\frm{#6}}

\newcommand{\rstick}[1]{*!L!<-.5em,0em>=<0em>{#1}}
\newcommand{\lstick}[1]{*!R!<.5em,0em>=<0em>{#1}}

\newcommand{\Qcircuit}{\xymatrix @*=<0em>}

\begin{document}

\global\long\def\U{\mathbb{U}}
\global\long\def\l({\left(}
\global\long\def\r){\right)}


\global\long\def\lc{\left\{}
\global\long\def\rc{\right\}}
\global\long\def\R#1{R_{\left|#1\right\rangle }}
\global\long\def\bra#1{\left\langle #1\right|}
\global\long\def\ket#1{\left|#1\right\rangle }
\global\long\def\op#1#2{\left|#1\right\rangle \left\langle #2\right|}
\global\long\def\ip#1#2{\left\langle #1,#2\right\rangle }
\global\long\def\s#1{\sum_{#1\in\left\{  0,1\right\}  ^{k}}}
\global\long\def\I{\mathbf{I}}
\global\long\def\ve{\varepsilon}

\newcommand{\eq}[1]{(\ref{eq:#1})}
\renewcommand{\sec}[1]{\hyperref[sec:#1]{Section~\ref*{sec:#1}}}
\newcommand{\fig}[1]{\hyperref[fig:#1]{Fig.~\ref*{fig:#1}}}
\newcommand{\tab}[1]{\hyperref[tab:#1]{Table~\ref*{tab:#1}}}
\newcommand{\routine}[1]{\hyperref[#1]{Routine~\ref*{#1}}}

\newcommand{\Gate}[1]{\textsc{#1}}
\newcommand{\hgate}{\Gate{h}}
\newcommand{\zgate}{\Gate{z}}
\newcommand{\sgate}{\Gate{s}}
\newcommand{\tgate}{\Gate{t}}
\newcommand{\pgate}{\Gate{p}}
\newcommand{\notgate}{\Gate{not}}
\newcommand{\cnotgate}{\Gate{cnot}}
\newcommand{\rzgate}{{\Gate{r}}_{z}}
\newcommand{\rygate}{{\Gate{r}}_{y}}

\newcommand{\GF}{\operatorname{GF}}
\DeclarePairedDelimiter\norm{\|}{\|}

\newcommand{\dm}[1]{\textbf{\color{green}[Dmitri: #1]}}
\newcommand{\ys}[1]{\textbf{\color{blue}[Yuan: #1]}}

\newtheorem{lemma}{Lemma}

\newcommand{\comment}[1]{}

\title{Approximate Quantum Fourier Transform with $O(n \log(n))$ T gates}

\author{Yunseong Nam}
\email{nam@ionq.co}
\affiliation{IonQ, College Park, MD 20740, USA}
\author{Yuan Su}
\email{buptsuyuan@gmail.com}
\affiliation{Department of Computer Science, Institute for Advanced Computer Studies, and Joint Center for Quantum
	Information and Computer Science, University of Maryland, College Park, MD 20740, USA}
\author{Dmitri Maslov}
\email{dmitri.maslov@gmail.com}
\affiliation{National Science Foundation, Alexandria, VA 22314, USA}
\altaffiliation{now with IBM Thomas J. Watson Research Center, Yorktown Heights, NY 10598, USA}

\date{\today}

\begin{abstract}
The ability to implement the Quantum Fourier Transform (QFT) efficiently on a quantum computer facilitates the advantages offered by a variety of fundamental quantum algorithms, such as those for integer factoring, computing discrete logarithm over Abelian groups, solving systems of linear equations, and phase estimation, to name a few.  The standard fault-tolerant implementation of an $n$-qubit unitary QFT approximates the desired transformation by removing small-angle controlled rotations and synthesizing the remaining ones into Clifford+$\tgate$ gates, incurring the $\tgate$-count complexity of $O(n \log^2(n))$.  In this paper, we show how to obtain approximate QFT with the $\tgate$-count of $O(n \log(n))$.  Our approach relies on quantum circuits with measurements and feedforward, and on reusing a special quantum state that induces the phase gradient transformation.  We report asymptotic analysis as well as concrete circuits, demonstrating significant advantages in both theory and practice.
\end{abstract}

\maketitle

\section{Introduction}
Quantum Fourier Transform (QFT) is one of the most important operations in quantum computing.  It can extract the periodicity encoded in the amplitudes of a quantum state, which is employed by an efficient algorithm for integer number factoring, widely known as Shor's algorithm \cite{ar:s}. Shor's integer factoring algorithm can be generalized (while still relying on the QFT) into a polynomial-time algorithm for the discrete logarithm problem over Abelian groups \cite{ar:s}. The importance of the above is witnessed through the threat such algorithms pose to modern public-key cryptosystems, such as the RSA or the ECC. Using the QFT as a subroutine, the eigenphase of a black-box unitary can be estimated up to an arbitrary precision \cite{ar:k}, which may be used to estimate quantum amplitudes \cite{co:bh, ar:g98}, simulate quantum chemistry/dynamics \cite{ar:kwp}, find the ground state/energy of a Hamiltonian \cite{ar:al}, compute Hessian to optimize molecular geometry \cite{ar:ka}, exponentiate unitaries \cite{ar:smm}, construct fractional powers of the QFT using constantly many copies of the controlled-QFT \cite{ar:smm, ar:kr}, extract features of the solution of linear systems \cite{ar:hhl}, and more.  QFT has also been used in quantum arithmetics \cite{ar:Draper,ar:RPGE} and quantum cryptography \cite{ar:YJSP}.

QFT can be implemented approximately by removing all rotation gates with angles smaller than a certain threshold value, resulting in the Approximate QFT (AQFT). In practice, it was shown that it suffices to apply AQFT with $\sim 5.3 \cdot 10^4$ controlled rotation gates to factor 2048-digit numbers (reflecting the de facto key size for today's standard \cite{rep:NIST}) with a high expected algorithmic accuracy ($\gtrsim 99.992\%$) \cite{ar:NB}. AQFT has been studied extensively in the literature. The robustness of the quantum computer equipped with the AQFT was investigated in detail \cite{ar:copper,ar:BEST,ar:NMI,ar:FH,ar:NB2}.  A study of the optimal level of the approximation of the AQFT in the presence of certain errors may be found in \cite{ar:NB3}.  Implementation of the QFT and its approximate version over restricted architectures was addressed in \cite{ar:tko, ar:m}.  An efficient approximate implementation of the AQFT that harnesses certain quantum hardware features was also investigated \cite{ar:MN}.

Quantum information is fragile, and it is generally accepted that the implementation of large quantum algorithms must rely on the fault-tolerant computations.  Fault tolerance suppresses the errors at the cost of using multiple physical qubits to encode a single logical qubit.  Fault-tolerant computations must furthermore rely on a quantum gate library consisting of those gates that are constructible fault tolerantly.  A standard choice for such a computationally universal gate library is Clifford+$\tgate$. Within known fault tolerance approaches, Clifford gates can generally be implemented with the relative ease, frequently transversally.  On the other hand, a non-Clifford gate typically does not admit such an implementation; for instance, a $\tgate$ gate may be implemented fault tolerantly by distilling a certain quantum state and then teleporting it into the gate \cite{ar:bk}.  A $\tgate$ gate is indeed far more costly than any of the Clifford gates, and therefore efficient fault-tolerant circuits must minimize the $\tgate$-count. 

To implement an $n$-qubit AQFT fault-tolerantly, the standard approach is to approximate the desired transformation by removing small-angle controlled rotations to bring down the gate count from $O(n^2)$ \cite[page 219]{bk:nc} to $O(n \log(n))$, and then replace the remaining $O(n \log(n))$ controlled rotations with their Clifford+$\tgate$ implementations. The resulting circuit has the $\tgate$-count of $O(n \log^2(n))$.  Only in the special case of the semiclassical version of AQFT \cite{ar:GN}, where the AQFT transform is followed by the measurement, the $\tgate$-count of $O(n \log(n))$ implementation is known \cite{ar:Goto}.  In contrast, in this paper, we focus on the fully coherent AQFT.

We develop a more efficient implementation with the $\tgate$-count complexity of $O(n \log(n))$ for the general case of fully coherent AQFT, improving over the standard construction by a factor of $O(\log(n))$.  Our results show that, in general and regardless of the amenability to the semiclassical approach, the AQFT may be implemented with $O(n \log(n))$ $\tgate$ gates.  This allows for the efficient implementation of the AQFT in any quantum algorithm, including those that use the AQFT as subroutines in the midst of the quantum computation; c.f. \cite{ar:kwp,ar:ka,ar:hhl,ar:RPGE,ar:YJSP}. Since our implementation is more involved compared to the standard, we also make a separate effort to show that the constant factor and small-order additive terms missing in the asymptotic analyses but otherwise present in our construction do not prevent it from achieving a significant practical advantage.

\section{Main result}
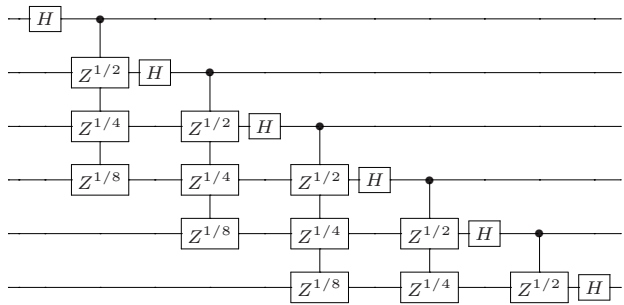
\begin{figure}[t]
\scriptsize
\vspace{-3.2mm}
\centering
	\begin{eqnarray*}
		\Qcircuit @C=0.5em @R=1.1em @!R{
			&\qw &\gate{H} &\ctrl{1}              &\qw      &\qw                   &\qw      &\qw                   &\qw      &\qw                   &\qw      &\qw            &\qw      &\qw\\
			&\qw &\qw      &\gate{Z^{1/2}}\qwx[1] &\gate{H} &\ctrl{1}              &\qw      &\qw                   &\qw      &\qw                   &\qw      &\qw            &\qw      &\qw\\
			&\qw &\qw      &\gate{Z^{1/4}}\qwx[1] &\qw      &\gate{Z^{1/2}}\qwx[1] &\gate{H} &\ctrl{1}              &\qw      &\qw                   &\qw      &\qw            &\qw      &\qw\\
			&\qw &\qw      &\gate{Z^{1/8}}        &\qw      &\gate{Z^{1/4}}\qwx[1] &\qw      &\gate{Z^{1/2}}\qwx[1] &\gate{H} &\ctrl{1}              &\qw      &\qw            &\qw      &\qw\\
			&\qw &\qw      &\qw                   &\qw      &\gate{Z^{1/8}}        &\qw      &\gate{Z^{1/4}}\qwx[1] &\qw      &\gate{Z^{1/2}}\qwx[1] &\gate{H} &\ctrl{1}       &\qw      &\qw\\
			&\qw &\qw      &\qw                   &\qw      &\qw                   &\qw      &\gate{Z^{1/8}}        &\qw      &\gate{Z^{1/4}}        &\qw      &\gate{Z^{1/2}} &\gate{H} &\qw
		}
	\end{eqnarray*}
	\vspace{-3.2mm}
	\caption{AQFT with $n=6$ and $b=3$. Note that each of the $n{-}1$ sets of controlled-$z^a$ gates are separated by the $\hgate$ gates.}
	\label{fig:AQFT}
\end{figure}

We start with an $n$-qubit AQFT whose construction relies on $O(nb)$ controlled-$\zgate^a$ gates with
\begin{equation*}
	\zgate^a:=
	\begin{bmatrix}
		1 & 0\\
		0 & e^{i\pi a}
	\end{bmatrix},
\end{equation*}
where $a \in \{1/2, 1/4, ..., 1/2^b\}$, for $b:=\lceil\log{n}\rceil$, and $n$ Hadamard ($\hgate$) gates (see \fig{AQFT} for an illustration with $n=6$ and $b=3$).  Such a choice of $b$ implies a very specific approximation error $\varepsilon$, whose analysis will be detailed in the next section. We unite the individual controlled rotations into $n{-}1$ sets separated by the $\hgate$ gates, such as illustrated in \fig{AQFT}. 

\comment{We drop the dependence on the algorithmic approximation error $\varepsilon$ in our analysis, noting that smaller errors $\varepsilon$ require only tiny adjustments to the value $b$ as outlined in \cite{ar:copper}. }

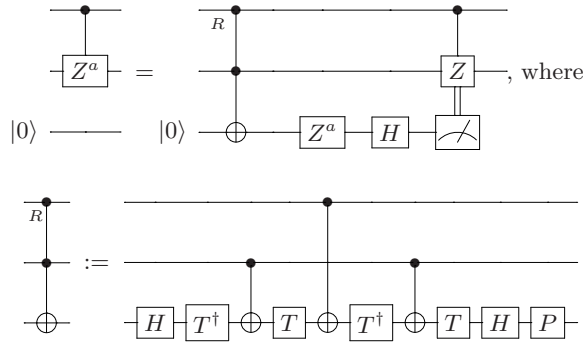
\begin{figure}[t]
\vspace{-3.2mm}
\begin{eqnarray*}
	\hspace{5mm}\Qcircuit @C=0.5em @R=1.2em @!R{
		&\ctrl{1} 		          &\qw\\
		&\gate{Z^a} &\qw\\
		\lstick{\ket{0}}	&\qw	     		          &\qw
	}
	&
	\raisebox{-2.55em}{\hspace{2mm}=\hspace{6mm}}
	&
	\Qcircuit @C=1.1em @R=1.1em @!R{
		&\ctrl{1}^R &\qw &\qw			&\qw			&\ctrl{1}		&\qw\\
		&\ctrl{1}     &\qw &\qw			&\qw			&\gate{Z} &\qw\\
		\lstick{\ket{0}}	&\targ        &\qw &\gate{Z^{a}}    &\gate{H}  &\meter \cwx[-1] 
	}
	\raisebox{-2.55em}{\text{, where}}
\end{eqnarray*}
\begin{eqnarray*}
	\Qcircuit @C=0.5em @R=1.65em @!R{
		&\ctrl{1}^R &\qw\\
		&\ctrl{1}   &\qw\\
		&\targ   	&\qw 
	}
	&
	\raisebox{-2.5em}{\hspace{2mm}:=\hspace{2mm}}
	&
	\Qcircuit @C=0.5em @R=1em @!R{
		& \qw 			& \qw		        & \qw		& \qw 		& \ctrl{2}	& \qw 		        & \qw 		& \qw 		& \qw 		& \qw      & \qw \\
		& \qw	 	    & \qw		        & \ctrl{1}	& \qw 		& \qw		& \qw 		        & \ctrl{1} 	& \qw 		& \qw 		& \qw      & \qw \\
		& \gate{H} 		& \gate{T^\dagger} 	& \targ 	& \gate{T}	& \targ		& \gate{T^\dagger} 	& \targ 	& \gate{T} 	& \gate{H} 	& \gate{P} & \qw 
	}
\end{eqnarray*}
\vspace{-3.2mm}
\caption{Ancilla-aided, measurement/feedforward-based fault-tolerant controlled-$\zgate^a$ gate.}
\label{fig:FTCZ}
\end{figure}

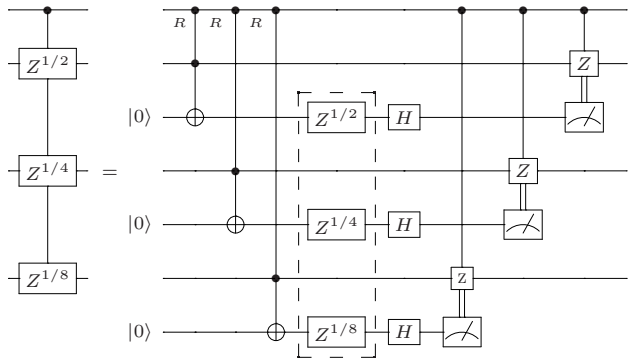
\begin{figure}[t]
\scriptsize
\vspace{-3.2mm}
	\begin{eqnarray*}
		\Qcircuit @C=0.5em @R=1.1em @!R{
			&\ctrl{1} 		        &\qw\\
			&\gate{Z^{1/2}}\qwx[2] &\qw\\
			& & \\
			&\gate{Z^{1/4}}\qwx[2] &\qw\\
			& & \\
			&\gate{Z^{1/8}}        &\qw
		}
		&
		\raisebox{-7.7em}{\hspace{2mm}=\hspace{6mm}}
		&
		\Qcircuit @C=1.1em @R=1.1em @!R{
			       &\ctrl{1}^R   &\ctrl{3}^R &\ctrl{5}^R &\qw	        &\qw			&\ctrl{5}        &\ctrl{3}        &\ctrl{1}		&\qw\\
			       &\ctrl{1}     &\qw        &\qw        &\qw		        &\qw			&\qw             &\qw             &\gate{Z} &\qw\\
\lstick{\ket{0}}   &\targ        &\qw        &\qw        &\gate{Z^{1/2}}  &\gate{H}  &\qw             &\qw             &\meter \cwx[-1] \\
                   &\qw          &\ctrl{1}   &\qw        &\qw                  &\qw            &\qw             &\gate{Z}   &\qw           &\qw\\
\lstick{\ket{0}}   &\qw          &\targ      &\qw        &\gate{Z^{1/4}}  &\gate{H}  &\qw             &\meter \cwx[-1]\\
                   &\qw          &\qw        &\ctrl{1}   &\qw                   &\qw            &\gate{\zgate}   &\qw             &\qw           &\qw\\
\lstick{\ket{0}}   &\qw          &\qw        &\targ      &\gate{Z^{1/8}}  &\gate{H}  &\meter \cwx[-1]
\gategroup{3}{5}{7}{5}{.9em}{--}
		}
	\end{eqnarray*}
\vspace{-3.2mm}
\caption{A 4-qubit example of the layer of controlled-$z^a$ gates. The uncontrolled rotations are grouped together to induce the phase gradient operation \cite{ar:g,bk:SKV}.}
\label{fig:crotlayer}
\end{figure}

To implement a given controlled-$\zgate^a$ rotation, we map its real-valued degree of freedom into that of the uncontrolled power of Pauli-$\zgate$, such as shown in \fig{FTCZ}.  This implementation was developed by combining Kitaev's trick \cite{ar:k} with Toffoli-measurement construction of Jones \cite{ar:j} with our own choice of the relative the phase Toffoli gate, and custom circuit simplifications.  Our circuit improves over the one reported in \cite[Fig.~10]{ar:ammr} (note that the middle $\tgate$ gate can be replaced with the $\zgate^a$ gate) by $4$ $\tgate$ gates ($8 \mapsto 4$), $9$ $\cnotgate$ gates ($12 \mapsto 3$), $1$ $\hgate$ gate ($4 \mapsto 3$), and $1$ Phase ($\pgate$) gate ($2 \mapsto 1$) at the cost of introducing $1$ measurement and $1$ classically-controlled controlled-$\zgate$ operation.  Note that the fault-tolerant cost of those operations introduced is significantly lower than that of a single $\tgate$ gate, as the construction of the $\tgate$ gate itself requires both a measurement and a classically controlled quantum correction \cite{ar:bk}.

\begin{figure*}[t]
\footnotesize
\vspace{-3.2mm}
	\begin{eqnarray*}
		\Qcircuit @C=1.1em @R=1.1em @!R{
\lstick{\ket{0}}               &{/}\qw &\gate{\Gate{$\psi$}}        &\multigate{1}{\Gate{Adder}_1} &\qw &\qw             &\qw &\qw &\qw                     &\qw    &\qw                     &\multigate{1}{\Gate{Adder}_2} &\qw &\qw            &\qw\\
\lstick{\ket{0}}               &{/}\qw &\multigate{1}{U_1}  &\ghost{\Gate{Adder}_1} &\gate{U'_1}        &\meter \cwx[1]  &    &    &\lstick{\ket{0}}        &{/}\qw &\multigate{1}{U_2} &\ghost{\Gate{Adder}_2}  &\gate{U'_2}      &\meter \cwx[1] &\rstick{\cdots}\\
\lstick{\ket{\psi_\text{in}}}  &{/}\qw &\ghost{U_1}                &\qw &\qw                       &\gate{U''_1} &\qw &\qw &\qw                     &\qw    &\ghost{U_2}               &\qw    &\qw                    &\gate{U''_2}&\qw
		}
\end{eqnarray*}
\vspace{-3.2mm}
\caption{A schematic diagram of the full implementation of the fault-tolerant AQFT. $\Gate{$\psi$}$ denotes the preparation of the special state $\ket{\psi_{b+1}}$.  $U_i$ illustrate the operations that precede the $i$th adder, including $\hgate$ gates and the relative phase Toffoli gates used to map controlled-$\zgate^a$ into uncontrolled $\zgate^a$ rotations. $U_i'$ denotes the operations that follow the adder up to the in-circuit measurements. $\Gate{Adder}_i$ denotes the $i$th adder. $U''_i$ are the classically controlled controlled-$\zgate$ gates, applied at the $i$th step.}
\label{fig:FTFT}
\end{figure*}
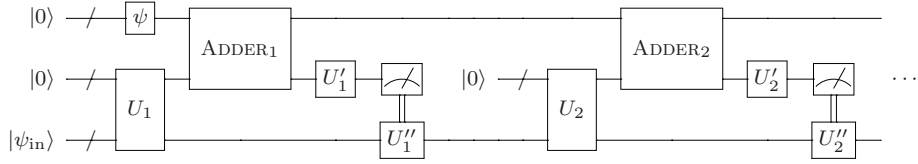

We now group the uncontrolled $\zgate^a$ rotations into one layer (time slice), as shown in \fig{crotlayer}.  This layer applies the transformation that was coined the phase gradient operation in \cite{ar:g}, the induction of which by the addition circuit was first reported in \cite{bk:SKV}.  Such a transformation can be implemented by a $b$-bit adder at the cost of $4b+O(1)$ $\tgate$ gates \cite{ar:g}, so long as one has access to a special quantum state $\ket{\psi_{b+1}}:=\frac{1}{\sqrt{2^{b+1}}}\sum_{j=0}^{2^{b+1}-1}e^{-2\pi i j/2^{b+1}}\ket{j}$.  The quantum state $\ket{\psi_{b+1}}$ can be reused to induce phase gradient transformations in all $n{-}1$ sets of controlled-$\zgate^a$ rotations. A schematic circuit diagram of our AQFT implementation is shown in \fig{FTFT}. 

To construct the special $(b+1)$-qubit state $\ket{\psi_{b+1}}$, we first apply $\hgate$ gates to the quantum register $\ket{00...0}$ and then exercise the gates $\zgate,\zgate^{-1/2}, ..., \zgate^{-1/2^{b}}$. The latter step is accomplished via approximating each $\zgate^a$ by RUS circuits \cite{ar:brs}. Specifically, we approximate complex number $e^{i\pi a}$ by $z^*/z$, where $z\in\mathbb{Z}[\omega]$ with $\omega:=e^{i\pi/4}$ being the cyclotomic integer obtained from the PSLQ Algorithm \cite{ar:b}. We choose $r\in\mathbb{Z}[\sqrt{2}]$ randomly and search the solution $y\in\mathbb{Z}[\omega]$ of the norm equation $|y|^2=2^L-|rz|^2$ with $L=\lceil \log(|rz|^2)\rceil$ \cite{ar:rs}, such that $V:=\frac{1}{\sqrt{2}^L}\Big(\begin{smallmatrix}rz & y\\-y^* & rz^*\end{smallmatrix}\Big)$ is a unitary. We exactly synthesize the two-qubit gate $\Big(\begin{smallmatrix}V & 0\\	0 & V^\dagger\end{smallmatrix}\Big)$ into a Clifford+$\tgate$ circuit \cite{ar:brs,ar:kmm}. Upon measuring the second qubit and obtaining $0$, the gate $\zgate^a$ is successfully implemented. Otherwise, a $\zgate$ error takes place and can be reversed at zero cost in the $\tgate$ gate count. The expected number of repetitions until success is $2^L/|rz|^2$. We resorted to using this more complex algorithm as opposed to the simpler one given by \cite{ar:kmm,ar:rs}, as we already use quantum circuits with measurements and feedforward elsewhere in our constructions, and the RUS approach results in about $2.5$-fold improvement \cite{ar:brs} in the number of the $\tgate$ gates required to obtain the desired $\zgate^a$.

\section{Details of the implementation}\label{sec:doi}
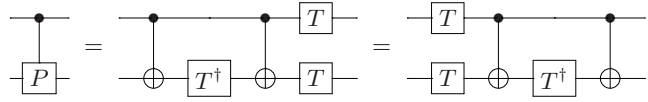
\begin{figure}[t]
\vspace{-3.2mm}
\[
\begin{array}{ccccc}
	\Qcircuit @C=0.5em @R=1.2em @!R{
		&\ctrl{1} 	   &\qw\\
		&\gate{P} &\qw\\
	}
	&
	\raisebox{-1.2em}{=}
	&
	\Qcircuit @C=1em @R=1em @!R{
	&\ctrl{1} &\qw     			 	&\ctrl{1}	&\gate{T} 	&\qw\\
	&\targ    &\gate{T^\dagger}		&\targ		&\gate{T}   &\qw \\
	}
	&
	\raisebox{-1.2em}{=}
	&
	\Qcircuit @C=1em @R=1em @!R{
	&\gate{T} 	&\ctrl{1} &\qw     			 	&\ctrl{1}	&\qw\\
	&\gate{T}   &\targ    &\gate{T^\dagger}		&\targ		&\qw \\
	}
\end{array}
\]
\vspace{-3.2mm}
\caption{Direct implementation of the controlled-$\pgate$ gate. These constructions also work when all $Z$-axis gates are replaced by their complex conjugates.}
\label{fig:CZa}
\end{figure}

In the following, we slightly optimize the improved fault-tolerant AQFT implementation described above.  We start by noting that the controlled-$\pgate$ gate may be implemented by two $\cnotgate$ gates and three $\tgate$ gates (including inverses) as shown in \fig{CZa}.  We know from our construction above that each controlled-$\zgate^a$ gate in the AQFT is implemented using 8 $\tgate$ gates.  Therefore, instead of relying on inducing the gradient operation through the adder, we implement controlled-$\pgate$ gates directly, according to \fig{CZa}.

Next, we consider controlled-$\tgate$ gates.  As per \fig{AQFT}, we see that each controlled-$\tgate$ gate in the AQFT neighbors a controlled-$\pgate$ gate in the following layer of controlled-$\zgate^a$ gates in the target qubit line. Since we implement controlled-$\pgate$ gates according to \fig{CZa}, we may obtain $\tgate$-count savings via gate cancellation ($\tgate\tgate^\dagger = Id$) by rewriting the controlled-$\tgate$ gate as the controlled-$\zgate^{3/4}$ gate followed by the controlled-$\zgate^{-1/2}$, where the controlled-$\zgate^{-1/2}$ gate is implemented according to \fig{CZa}, inducing $\tgate$-count reduction by $2$ on the `target' of controlled-$\zgate^{-1/2}$ and controlled-$\tgate$ gates, and by another $2$ for each layer of controlled-$\zgate^a$ gates by cancellations on the `control' line, and the controlled-$\zgate^{3/4}$ gate is implemented directly as per the top panel of \fig{FTCZ}, which costs $5$ $\tgate$ gates.

Altogether, the above implementation of the controlled-$\tgate$ and controlled-$\pgate$ gate pair requires $7 (=5{+}3{+}3{-}2{-}2)$ $\tgate$ gates.  This is in comparison to $16$ $\tgate$ gates that would otherwise have been used by the implementation based on the adder.  What remains to be investigated at this point is the modification that needs to be made to the gradient operation so as to induce a partial gradient operation, i.e., $\ket{k}\ket{\psi_{d+1,b+1}} \mapsto e^{2\pi ik/2^{b+1}}\ket{k}\ket{\psi_{d+1,b+1}}$, where $k<2^{b-d}$, $d \leq b$, and $\ket{\psi_{d+1,b+1}}$ is the state $\ket{\psi_{b+1}}$ without first $d{+}1$ qubits, to implement the remaining $\zgate^a$ gates in a layer.

To obtain the partial gradient operation, we analyze how the gradient operation works.  Firstly, we formally define the state $\ket{\psi_{d+1,b+1}} :=  \frac{1}{\sqrt{2^{b-d}}} \sum_{j=0}^{2^{b-d}-1}e^{-2\pi i j/2^{b+1}}\ket{j}$.  The application of $(b{-}d)$-bit addition (see \cite{ar:g}) to $\ket{k}\ket{\psi_{d+1,b+1}}$ results in two cases: $k{+}j < 2^{b-d}$ and $k{+}j \geq 2^{b-d}$.  In order for the partial gradient operation to work, we need $k{+}j \mapsto k{+}j \mod 2^{b-d}$.  This may be achieved by applying $\zgate^{1/2^d}$ gate to the most significant bit of the modular addition circuit.  Since in our case $d = 2$, this amounts to applying a $\tgate$ gate for each gradient operation.  This means that the overall result of our optimization detailed in this section is by about $8(n-2)$ $\tgate$ gates.

\section{Comparisons to prior work}
Our improved implementation of AQFT${}_n$ with $n\,{>}\,b\,{>}\,2$ requires the qubit count of $n_q = n+3b-4$, the $\cnotgate$-gate count of $7.5n\,{-}\,13 + \sum_{l=3}^{n-1} (16\min(b\,{-}\,2,l\,{-}\,2)\,{-}\,5) + \sum_{b'=3}^{\min({b,n-1})} C_{\cnotgate}({\rm RUS}_{b'})/p_{b'}$, and the $\tgate$-count of $7n-11 + \sum_{l=3}^{n-1}(8\min(b-2,l-2)+1)+ \sum_{b'=3}^{\min({b,n-1})} C_{\tgate}({\rm RUS}_{b'})/p_{b'}$, where $C_{g}({\rm RUS}_{b'})$ denotes the count of the fault-tolerant gate $g$ in the RUS circuit synthesizing $z^{-1/2^{b'}}$, and $p_{b'}$ denotes the success probability of the RUS circuit.  As follows from our constructions, the $\tgate$ gate count can be fairly accurately approximated by the simple formula $8n(b{-}1)$.  This may be compared to the previous state of the art that uses a variant of \cite[Fig.~10]{ar:ammr} to implement the controlled-$\zgate^a$, which requires $n_q{=}n{+}1$ qubits, the $\cnotgate$ gate count of $12\cdot \sum_{l=0}^{n-1}\min(b,l)$, and the $\tgate$-count of $3(n-1) + \sum_{b'=2}^{\min(b,n-1)} (n-b') [C_{\tgate}({\rm Gridsynth}_{b'})+8]$, where $C_{\tgate}({\rm Gridsynth})$ is the $\tgate$-count of the Gridsynth algorithm \cite{ar:rs} synthesizing $z^{1/2^{b'}}$ and $C_{\tgate} = 1$ when considering $z^{\pm1/4}$ gate.

For a concrete comparison with the previous state of the art \cite{ar:nrscm,ar:ammr} at the gate-by-gate level, we implemented our improved fault-tolerant construction as described in \sec{doi} in software.  We synthesized the RUS circuits for $z^a$ gates with $a \in \{-1/2^3, -1/2^4, ..., -1/2^{13}\}$, motivating the choice of the smallest angle $\pi/2^{b}$ by that sufficient to launch a quantum attack on the classically-infeasible instance of the integer factoring problem corresponding to cracking the RSA-2048.  We also chose the overall fault-tolerance error that arises from the gate synthesis to be below $1.1 \cdot 10^{-4}$ for all sizes of the AQFT ($n \leq 4096$ and $b = 13$) we considered.  In particular, we chose the error $10^{-5}$ per $z^a$ gate approximation for our improved construction. This amounts to the gate-synthesis error budget of $\sim 10^{-5}/n$ per rotation for the previous state-of-the-art AQFT circuit.  The improvement of the accuracy per $\zgate^a$ gate is justified by the fact that our implementation of the AQFT requires the approximation of only $O(b)$ rotations instead of $O(nb)$ in the previous constructions.

Summary of the resulting quantum resource cost of our improved AQFT implementation is shown in \tab{comparison}.  We included a comparison of the gate costs of our implementation to those circuits known previously: first set relying on \cite[Fig.~10]{ar:ammr} to implement controlled-$\zgate^a$ gates in the AQFT and the second set resulting from an automated AQFT circuit optimization \cite{ar:nrscm}.  For both implementations, we used Gridsynth algorithm \cite{ar:rs} to synthesize $\zgate^a$ gates.  Note that our implementation carries a significant practical advantage, saving quantum resource cost in the form of the $\tgate$-count by a factor of as large as $12$ (AQFT$_{4096}$ with $b=13$). The slight increase in $n_q$ and the $\cnotgate$-gate counts are completely offset by the savings in the $\tgate$-count in the fault-tolerant regime.

\begin{table*}
	\caption{Quantum resource counts for implementing an $n$-qubit AQFT with $b=13$.  $n_q$ denotes the number of qubits required to execute the corresponding circuit.  Columns $\cnotgate$ and $\tgate$ report the number of respective gates in the circuits. All circuits are available in \cite{github}.}
	\label{tab:comparison}
	\begin{tabular}{|c||c|c|c||c|c|c||c|c|c|} 
		\hline
		&\multicolumn{3}{c||}{Our AQFT implementation}
		&\multicolumn{3}{c||}{AQFT with controlled-$\zgate^a$ per \cite[Fig.~10]{ar:ammr}} 
		&\multicolumn{3}{c|}{Optimized AQFT \cite{ar:nrscm}} \\ \cline{1-10}
		Circuit & $n_{q}$ & $\cnotgate$ & $\tgate$ & $n_q$ & $\cnotgate$ & $\tgate$ 
		& $n_q$ & $\cnotgate$ & $\tgate$ \\ \hline \hline
		AQFT$_{8}$ & 25 & 390 & 303 & 9 & 336 & 1,083 & 8 & 56 & 1,821 \\ \hline
		AQFT$_{16}$ & 51 & 1,798 & 1,162 & 17 & 1,404 & 6,309 & 16 & 234 & 7,815 \\ \hline
		AQFT$_{32}$ & 67 & 4,654 & 2,698 & 33 & 3,900 & 19,261 & 32 & 650 & 22,683 \\ \hline
		AQFT$_{64}$ & 99 & 10,366 & 5,770 & 65 & 8,892 & 47,099 & 64 & 1,482 & 54,269 \\ \hline
		AQFT$_{128}$ & 163 & 21,790 & 11,914 & 129 & 18,876 & 106,631 & 128 & 3,146 & 123,333 \\ \hline
		AQFT$_{256}$ & 291 & 44,638 & 24,202 & 257 & 38,844 & 229,729 & 256 & 6,474 & 267,007 \\ \hline
		AQFT$_{512}$ & 547 & 90,334 & 48,778 & 513 & 78,780 & 476,873 & 512 & 13,130 & 553,277 \\ \hline
		AQFT$_{1024}$ & 1,059 & 181,726 & 97,930 & 1,025 & 158,652 & 993,727 & 1,024 & 26,442 & 1,148,497 \\ \hline
		AQFT$_{2048}$ & 2,083 & 364,510 & 196,234 & 2,049 & 318,396 & 2,084,983 & 2,048 & 53,066 & 2,427,081 \\ \hline
		AQFT$_{4096}$ & 4,131 & 730,078 & 392,842 & 4,097 & 637,884 & 4,316,993 & 4,096 & 106,314 & 4,993,035 \\ \hline 
		\hline
		Complexity & $O(n)$ & $O(n\log(n))$ & $O(n\log(n))$ & $O(n)$ & $O(n\log(n))$ & $O(n\log^2(n))$ 
		& $O(n)$ & $O(n\log(n))$ & $O(n\log^2(n))$  \\ \hline
	\end{tabular}
	\vspace{-3.2mm}
\end{table*}

\section{Complexity analysis}
The total $\tgate$-count in our AQFT circuit is $8n(b-1)+O(b\log(b/\varepsilon))$. This is because each of the $nb{-}b(b+1)/2=nb+O(b^2)$ controlled-$\zgate^a$ gates consumes $4$ $\tgate$ gates to be first mapped into an uncontrolled $\zgate^a$ and another $4$ $\tgate$ gates for the $\zgate^a$ to be implemented as a part of the adder circuit, except for controlled-$\zgate^{1/2}$ and controlled-$\zgate^{1/4}$ gates; the two require 7 $\tgate$ gates to implement and 1 $\tgate$ gate to correct for the  phase in the partial gradient operation. The construction of the special state $\ket{\psi_{d+1,b+1}}$ requires implementation of $O(b)$ $\zgate^a$ rotations, and we approximate each rotation with $O(\log(b/\varepsilon))$ $\tgate$ gates \cite{ar:brs} to achieve accuracy $\varepsilon/b$ per rotation. 

There are two sources of approximation errors in our construction. Our circuit differs from the ideal AQFT circuit only in the preparation of the special state $\ket{\psi_{d+1,b+1}}$. Therefore, the spectral norm distance between our AQFT circuit and the ideal AQFT is $O(b\cdot\varepsilon/b)=O(\varepsilon)$. This ensures that, with $1-O(\varepsilon^2)$ probability, regardless of how many operations to follow from the $\ket{\psi_{d+1,b+1}}$ state preparation stage, our circuit implements the ideal AQFT. If we choose $b=O(\log (n/\varepsilon))$, the spectral norm error of the ideal AQFT circuit will be $O(\varepsilon)$. Due to the triangle inequality, the total error can be upper bounded by adding the error of the Clifford+$\tgate$ synthesis and the error of AQFT, which is still $O(\varepsilon)$.

The above error analysis shows that for all effective purposes (specifically, when $\varepsilon \succ n/2^n$) we can drop the dependence on the approximation error $\varepsilon$, resulting in the claimed $\tgate$-count of $O(n\log n)$.

\section{Conclusion}
Before our contribution, the best known coherent approximation of the $n$-qubit QFT to an error $\varepsilon$ by a quantum fault-tolerant Clifford+$\tgate$ circuit featured the $\tgate$-count of $O\left(n\log(n/\varepsilon)\log(\frac{n\log(n/\varepsilon)}{\varepsilon})\right)$, with the term $O(n\log(n/\varepsilon))$ originating from the standard AQFT construction using controlled rotations, and term $O\left(\log(\frac{n\log(n/\varepsilon)}{\varepsilon})\right)$ coming from the fault-tolerance overhead.  In this paper we reported an improved approximation of the QFT by a quantum Clifford+$\tgate$ circuit with the $\tgate$-count of $O\left(n\log(n/\varepsilon)+\log(n/\varepsilon)\log(\frac{\log(n/\varepsilon)}{\varepsilon})\right)$.  Our improvement is twofold: first, we reduce the dependence on $n$ from $O(n\log^2(n))$ to $O(n\log(n))$, and second, we moved the dependence on $\varepsilon$ from the leading term into a lower order additive term.  This means that the smaller the desired approximation error the more efficient our construction is compared to those known previously. 

Our implementation includes constant factor improvements that are not captured by the asymptotics.  We report significant practical advantages from applying our construction, as is evidenced by the numbers in \tab{comparison}, showing the improvement by a factor of $10$ to $12$ in the $\tgate$-count for values of $n$ of the size expected in practical applications of quantum computers.  This shows that our result carries both theoretical and practical value.

\section*{Acknowledgments}
YS thanks Andrew Glaudell for helpful discussions.  Authors thank Craig Gidney for his helpful comments that inspired us to further improve our results.  We note that Craig Gidney independently developed $6n+Const$ improvement through the use of reduced gradient operation ($8n+Const$ in our version).  

YS was supported in part by the Army Research Office (MURI award W911NF-16-1-0349) and the National Science Foundation (grant 1526380).  This material was based on work supported by the National Science Foundation, while DM working at the Foundation.  Any opinion, finding, and conclusions or recommendations expressed in this material are those of the author and do not necessarily reflect the views of the National Science Foundation.


\begin{thebibliography}{10}

\bibitem{ar:s}
P. Shor,
\newblock Polynomial-time algorithms for prime factorization and discrete logarithms on a quantum computer,
\newblock SIAM J. Comput. {\bf 26}, 1484--1509 (1997), 
\newblock \href{https://arxiv.org/abs/quant-ph/9508027}{arXiv:quant-ph/9508027}.

\bibitem{ar:k}
A. Kitaev, 
\newblock Quantum measurements and the {A}belian {S}tabilizer {P}roblem,  
\newblock (1995), \href{https://arxiv.org/abs/quant-ph/9511026}{arXiv:quant-ph/9511026}.

\bibitem{co:bh}
G. Brassard and P. Hoyer,
\newblock An exact quantum polynomial-time slgorithm for {S}imon's Problem,
\newblock Proc. of Fifth Israeli Symposium on Theory of Computing and Systems, pages 12--23 (1997),
\newblock \href{https://arxiv.org/abs/quant-ph/9704027}{arXiv:quant-ph/9704027}.

\bibitem{ar:g98}
L. Grover,
\newblock Quantum computers can search rapidly by using almost any transformation,
\newblock Phys. Rev. Lett. {\bf 80}, 4329 (1998),
\newblock \href{https://arxiv.org/abs/quant-ph/9712011}{arXiv:quant-ph/9712011}.



\bibitem{ar:kwp}
I. Kassal, J.~D. Whitfield, A. Perdomo-Ortiz, M.-H. Yung, and A. Aspuru-Guzik,
\newblock Simulating chemistry using quantum computers,
\newblock Annu. Rev. Phys. Chem. {\bf 62}, 185 (2011),
\newblock \href{https://arxiv.org/abs/1007.2648}{arXiv:1007.2648}.


\bibitem{ar:al}
D.~S. Abrams and S. Lloyd,
\newblock Quantum algorithm providing exponential speed increase for finding eigenvalues and eigenvectors,
\newblock Phys. Rev. Lett. {\bf 83}, 5162 (1999), \href{https://arxiv.org/abs/quant-ph/9807070}{arXiv:quant-ph/9807070}.

\bibitem{ar:ka}
I. Kassal and A. Aspuru-Guzik,
\newblock Quantum algorithm for molecular properties and geometry optimization,
\newblock J. Chem. Phys. {\bf 131}, 224102 (2009),
\newblock \href{https://arxiv.org/abs/0908.1921}{arXiv:0908.1921}.

\bibitem{ar:smm}
L. Sheridan, D. Maslov, and M. Mosca,
\newblock Approximating fractional time quantum evolution,
\newblock J. Phys. A: Math. Theor. {\bf 42}, 185302 (2009),
\newblock \href{https://arxiv.org/abs/0810.3843}{arXiv:0810.3843}.

\bibitem{ar:kr}
A. Klappenecker and M. Roetteler,
\newblock Quantum software reusability,
\newblock Int. J. Foundations Computer Science {\bf 14}(5), 777--796 (2003),
\newblock \href{https://arxiv.org/abs/quant-ph/0309121}{arXiv:quant-ph/0309121}.

\bibitem{ar:hhl}
A.~W. Harrow, A. Hassidim, and S. Lloyd,
\newblock Quantum algorithm for linear systems of equations,
\newblock Phys. Rev. Lett. {\bf 103}, 150502 (2009), 
\newblock \href{https://arxiv.org/abs/0811.3171}{arXiv:0811.3171}.

\bibitem{ar:Draper}
T.~G. Draper,
\newblock Addition on a quantum computer,
\newblock (2000), \href{https://arxiv.org/abs/quant-ph/0008033}{arXiv:quant-ph/0008033}.

\bibitem{ar:RPGE}
L. Ruiz-Perez and J.~C. Garcia-Escartin,
\newblock Quantum arithmetic with the quantum {F}ourier transform,
\newblock Quantum Inf. Process. {\bf 16}, 152 (2017),
\newblock \href{https://arxiv.org/abs/1411.5949}{arXiv:1411.5949}.

\bibitem{ar:YJSP}
Y.--G. Yang, X. Jia, S.--J. Sun, and Q.--X. Pan,
\newblock Quantum cryptographic algorithm for color images using quantum {F}ourier transform and double random-phase encoding,
\newblock Inf. Sci. {\bf 277}, 445--457 (2014).

\bibitem{rep:NIST}
E. Barker and A. Roginsky,
\newblock NIST Speical Publication 800-131A Revision 1
\newblock (NIST, Gaithersburg, MD, 2015),
\newblock \href{https://csrc.nist.gov/publications/detail/sp/800-131a/rev-1/final}{SP 800-131A Rev. 1}.

\bibitem{ar:NB}
Y.~S. Nam and R. Bl\"umel,
\newblock Scaling laws for Shor's algorithm with a banded quantum {F}ourier transform,
\newblock Phys. Rev. A {\bf 87}, 032333 (2013).



\bibitem{ar:copper}
D. Coppersmith, (2002), 
\newblock An approximate Fourier transform useful in quantum factoring,
\newblock \href{https://arxiv.org/abs/quant-ph/0201067}{arXiv:quant-ph/0201067}.

\bibitem{ar:BEST}
A. Barenco, A. Ekert, K.-A. Suominen, and P. T\"orm\"a,
\newblock Approximate quantum Fourier transform and decoherence,
\newblock Phys. Rev. A {\bf 54}, 139 (1996),
\newblock \href{https://arxiv.org/abs/quant-ph/9601018}{quant-ph/9601018}.

\bibitem{ar:NMI}
J. Niwa, K. Matsumoto, and H. Imai,
\newblock General-purpose parallel simulator for quantum computing,
\newblock Phys. Rev. A {\bf 66}, 062317 (2002),
\newblock \href{https://arxiv.org/abs/quant-ph/0201042}{arXiv:quant-ph/0201042}.

\bibitem{ar:FH}
A. Fowler and L.~C.~L. Hollenberg,
\newblock Scalability of Shor's algorithm with a limited set of rotation gates,
\newblock Phys. Rev. A {\bf 70}, 032329 (2004),
\newblock \href{https://arxiv.org/abs/quant-ph/0306018}{arXiv:quant-ph/0306018}.


\bibitem{ar:NB2}
Y.~S. Nam and R. Bl\"umel,
\newblock Robustness and performance scaling of a quantum computer with respect to a class of static defects,
\newblock Phys. Rev. A {\bf 88}, 062310 (2013).

\bibitem{ar:NB3}
Y.~S. Nam and R. Bl\"umel,
\newblock Analytical formulas for the performance scaling of quantum processors with a large number of defective gates,
\newblock Phys. Rev. A {\bf 92}, 042301 (2015).

\bibitem{ar:tko}
Y. Takahashi, N. Kunihiro, and K. Ohta,
\newblock The quantum Fourier transform on a linear nearest neighbor architecture,	
\newblock Quantum Inf. Comput. {\bf 7}(4), 383--391 (2007).

\bibitem{ar:m}
D. Maslov,
\newblock Linear depth stabilizer and quantum Fourier transformation circuits with no auxiliary qubits in finite neighbor quantum architectures,
\newblock Phys. Rev. A {\bf 76}, 052310 (2007),
\newblock \href{https://arxiv.org/abs/quant-ph/0703211}{arXiv:quant-ph/0703211}.

\bibitem{ar:MN}
D. Maslov and Y.~S. Nam,
\newblock Use of global interactions in efficient quantum circuit constructions,
\newblock New J. Phys. {\bf 20}, 033018 (2018),
\newblock \href{https://arxiv.org/abs/1707.06356}{arXiv:1707.06356}.

\bibitem{ar:bk}
S. Bravyi and A. Kitaev, 
\newblock Universal quantum computation with ideal {C}lifford gates and noisy ancillas,
\newblock Phys. Rev. A {\bf 71}, 022316 (2005),
\newblock \href{https://arxiv.org/abs/quant-ph/0403025}{quant-ph/0403025}.

\bibitem{bk:nc}
M.~A. Nielsen and I.~L. Chuang, 
\newblock {\em Quantum Computation and Quantum Information},
\newblock Cambridge University Press, New York (2000).

\bibitem{ar:GN}
R.~B. Griffiths and C.~S. Niu,
\newblock Semiclassical {F}ourier transform for quantum computation,
\newblock Phys. Rev. Lett. {\bf 76}, 3228 (1996).
\newblock \href{https://arxiv.org/abs/quant-ph/9511007}{quant-ph/9511007}.

\bibitem{ar:Goto}
H. Goto,
\newblock Resource requirements for a fault-tolerant quantum {F}ourier transform,
\newblock Phys. Rev. A {\bf 90}, 052318 (2014),
\newblock \href{https://arxiv.org/abs/1410.5124}{arXiv:1410.5124}.

\bibitem{ar:j}
C. Jones, 
\newblock Novel constructions for the fault-tolerant {T}offoli gate,
\newblock Phys. Rev. A {\bf 87}, 022328 (2013),
\newblock \href{https://arxiv.org/abs/1212.5069}{arXiv:1212.5069}.


\bibitem{ar:ammr}
M. Amy, D. Maslov, M. Mosca, and M. Roetteler,
\newblock A meet-in-the-middle algorithm for fast synthesis of depth-optimal quantum circuits,
\newblock IEEE Trans. CAD {\bf 32}(6), 818--830 (2013),
\newblock \href{https://arxiv.org/abs/1206.0758}{arXiv:1206.0758}.

\bibitem{ar:g}
C. Gidney,
\newblock Halving the cost of quantum addition,
\newblock Quantum {\bf 2}, 74 (2018),
\newblock \href{https://arxiv.org/abs/1709.06648}{arXiv:1709.06648}.

\bibitem{bk:SKV}
A.~Yu. Kitaev, A.~H. Shen, and M.~N. Vyalyi, 
\newblock {\em Classical and Quantum Computation},
\newblock American Mathematical Society, Providence, RI (2002).


\bibitem{ar:brs}
\newblock A. Bocharov, M. Roetteler, and K.~M. Svore,
\newblock Efficient synthesis of universal Repeat-Until-Success circuits,
\newblock  Phys. Rev. Lett. {\bf 114}, 080502 (2015), 
\newblock \href{https://arxiv.org/abs/1404.5320}{arXiv:1404.5320}.

\bibitem{ar:b}
P. Bertok, (2004), \newline
\newblock PSLQ integer relation algorithm implementation, (2004),
\newblock \href{http://library.wolfram.com/infocenter/MathSource/4263/}{http://library.wolfram.com/infocenter/MathSource/4263/}.

\bibitem{ar:rs}
N.~J. Ross and P. Selinger,
\newblock Optimal ancilla-free {C}lifford+{T} approximation of z-rotations,
\newblock Quantum Inf. Comput. {\bf 16}(11-12), 901--953 (2016), 
\newblock \href{https://arxiv.org/abs/1403.2975}{arXiv:1403.2975}.

\bibitem{ar:kmm}
V. Kliuchnikov, D. Maslov, and M. Mosca, 
\newblock Fast and efficient exact synthesis of single-qubit unitaries generated by {C}lifford and {T} gates, 
\newblock Quantum Inf. Comput. {\bf 13}(6-7), 607--630 (2013), 
\newblock \href{https://arxiv.org/abs/1206.5236}{arXiv:1206.5236}.

\bibitem{ar:nrscm}
Y.~S. Nam, N.~J. Ross, Y. Su, A.~M. Childs, and D. Maslov,
\newblock Automated optimization of large quantum circuits with continuous parameters,
\newblock {\em npj} Quantum Inf. {\bf 4}, 23 (2018),
\newblock \href{https://arxiv.org/abs/1710.07345}{arXiv:1710.07345}.

\bibitem{github}
Y.~S. Nam, Y. Su, and D. Maslov, 
\newblock (2018), \href{https://github.com/y-nam/QFT}{https://github.com/y-nam/QFT}.
\end{thebibliography}

\end{document}